\documentclass[paper]{JINST} % 10pt is ignored!

\usepackage{epsfig,multicol,bbm}

\title{The ATLAS High Level Trigger Region of Interest Builder}

\author{Robert Blair, John Dawson, Gary Drake, William Haberichter, James Schlereth, Jinlong Zhang \\
Argonne National Laboratory, 9700 S. Casse Ave., Argonne, IL 60439, USA\\
        E-mail: \email{reb@anl.gov}, \email{jwd@hep.anl.gov}, \email{drake@hep.anl.gov}, \email{wnh@anl.gov}, \email{jls@hep.anl.gov},  \email{zhangjl@anl.gov}}
\author{Maris Abolins, Yuri Ermoline, Bernard Pope\\
Michigan State University, East Lansing, MI 48824, USA\\
	E-mail:  \email{abolins@pa.msu.edu}, \email{Yuri.Ermoline@cern.ch}, \email{pope@pa.msu.edu}
}

\preprint{ATL-DAQ-PUB-2007-001}	% OR: \preprint{Aaaa/Mm/Yy\\Aaa-aa/Nnnnnn}
\abstract{This article describes the design, testing and production of the
ATLAS Region of Interest Builder (RoIB).  This device acts as an interface between the
Level 1 trigger and the high level trigger (HLT) farm for the ATLAS LHC detector.
It distributes all of the level 1 data for a subset
of events to a small number of (16 or less) individual commodity processors.  These processors in turn
provide this information to the HLT.  This allows the HLT to use the level 1 information to narrow data requests to
areas of the detector where level 1 has identified interesting objects. }


\begin{document} 

%\maketitle  IS IGNORED %%%%%%%%%%%

\section{Introduction}
The ATLAS trigger has a hardware based Level 1 system to make an early event selection which determines whether there are any objects of interest above programmable thresholds\cite{AtlasLVL1}.  The remaining levels of the ATLAS trigger use information from the hardware based Level 1 system to guide the retrieval of information from the readout system\cite{AtlasHLT}.  Jet, electromagnetic and tau clusters, missing Et, total Et, total jet Et and muon candidate information from Level 1 determine Regions of Interest (RoIs) that seed further trigger decisions.  The design of the Level 1 trigger and the front end electronics allows the rate of Level 1 accepts, L1A, to be as high as 100kHz.  Once a Level 1 decision is made, all the data from the detector is read out into buffers which can be interrogated via gigabit ethernet links.  The remaining part of the trigger decision is conducted using commodity processors and the data is read with commodity ethernet networking.  To keep the data volume low the High Level Trigger (HLT), defined as the stages which follow Level 1, uses the detailed information from Level 1 to guide data requests.  Making these data available to the HLT requires collecting information from a number of components of the Level 1 system.  This collection has to occur at the full Level 1 accept rate and can not be achieved with a single commodity processor.  The ATLAS Region of Interest Builder (RoIB) builds the various pieces of information from the Level 1 system into complete event records.  It then passes these data to one or more Level 2 Supervisors (L2SV) which are commodity processors that distribute this data to the HLT via the network.  Each L2SV will receive all the data for a subset of the events and pick which processor will make a decision for each of the events it receives and pass the Level 1 data on to the processor that will run the trigger algorithms.  In this way no single L2SV has to deal with the full rate of Level 1.  The rate to any one L2SV is divided by the number of L2SV's in the system.

\FIGURE[l]{\epsfig{file=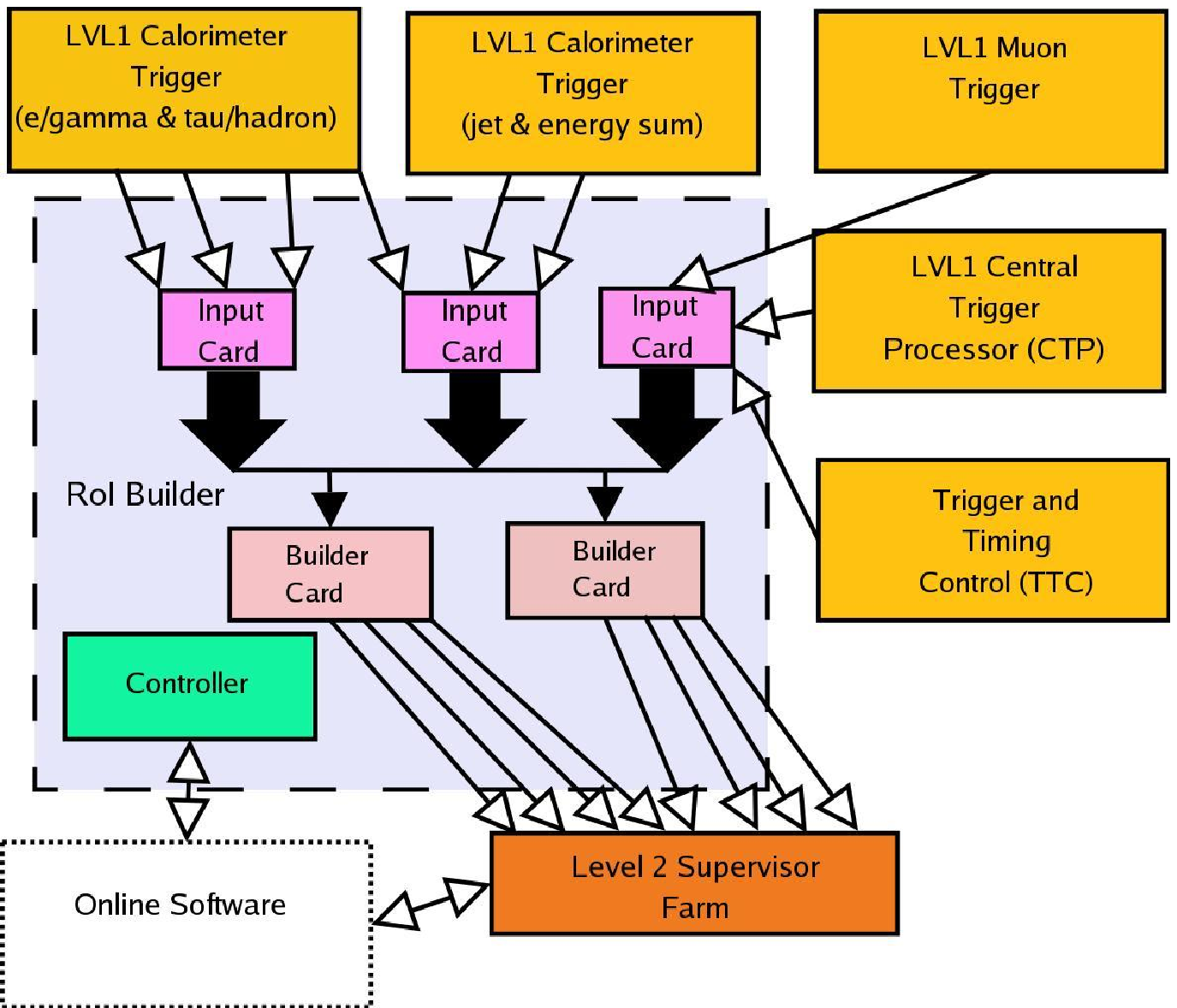,width=10cm} 
        \caption[Region of Interest Builder system]{This figure shows the context of the RoIB and its components.  A system with three input and two builder cards is depicted.}%
	\label{LVL1-RoIB}}

\FIGURE{\epsfig{file=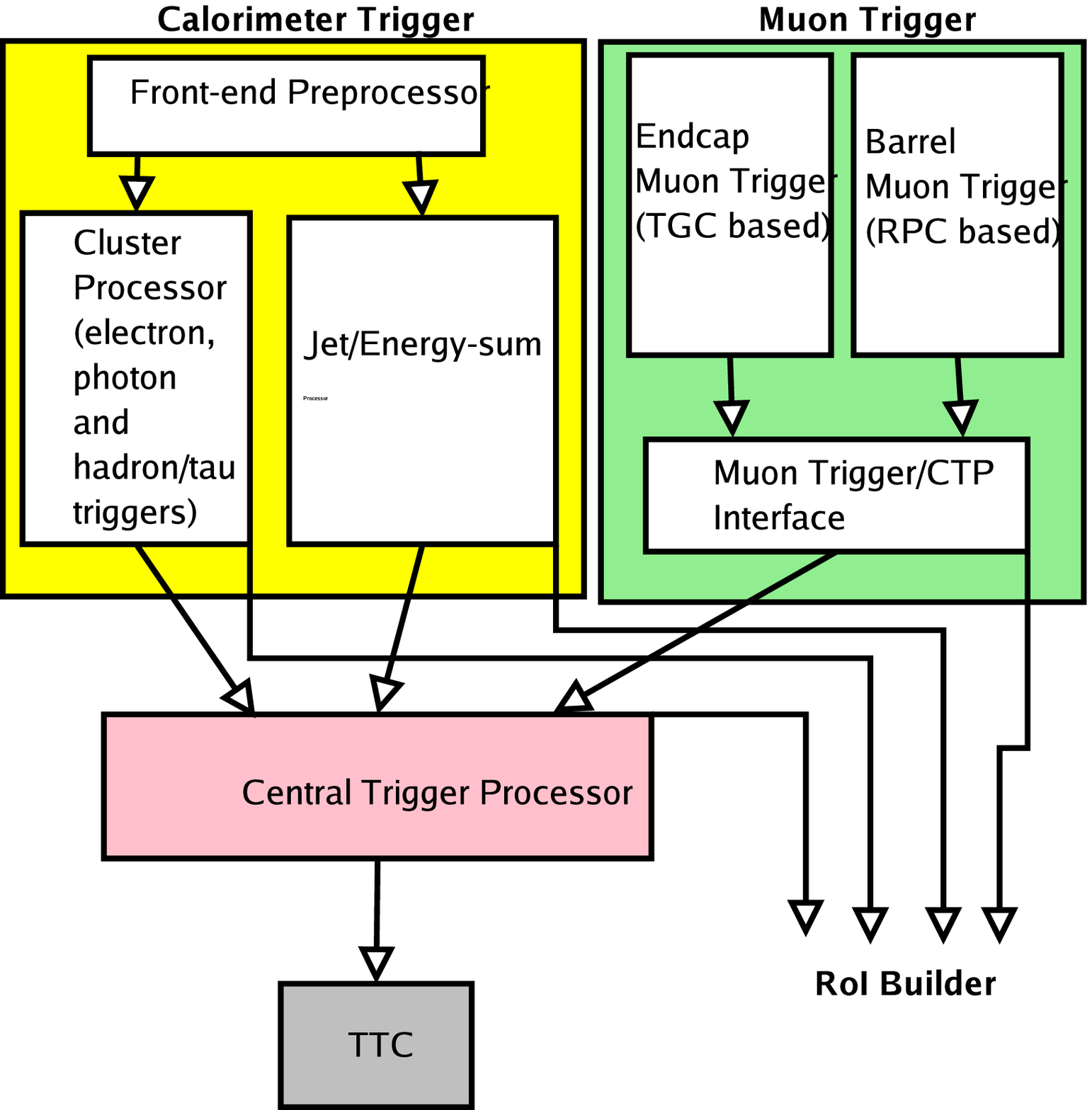,width=10cm} 
        \caption[ATLAS Level 1]{This figure shows the components of Level 1 that communicate with the RoIB.  The lines from the Level 1 components represent one or more S-link connections (see table~\ref{LVL1-data} for more details on the data transferred and the number of links).}%
	\label{L1}}

\TABLE{\begin{tabular}{|c|c|l|} \hline
{\bf Level 1 system} & {\bf number of links} & {\bf data transferred} \\ \hline
 \hline
Central Trigger Processor & 1 & input state (160 bits), trigger decision (256 bits),\\
& &   trigger type (8 bits),GPS time (32 bits),\\
& &  internal trigger data (32 bits) \\ 
\hline Muon System & 1 & list of up to 16 $P_T$ ordered muon candidates \\
& & including threshold passed and \\ 
& & location of the candidate (32 bits each) \\  
\hline Calorimeter ($e,\gamma,\tau$) & 4 & Thresholds passed (16 bits), saturation flag (1 bit), \\ 
& & position data (12 bits) for each trigger entity\\ 
\hline Calorimeter (jet and energy sum) & 2 & Thresholds passed (12 bits), saturation flag (1 bit), \\
& & position data (10 bits) for each jet \\ 
 & & $E_x$ (16 bits), $E_y$ (16 bits), $\Sigma E_T$ thresholds passed (4 bits),\\
 & & $\Sigma E_T$ (16 bits), missing $E_T$ thresholds passed (8 bits),\\
& &  jet $E_T$ sum thresholds passed (4bits) \\ 
\hline Trigger and Timing Control (TTC) & TTC fiber & Trigger Type (8 bits), Bunch Counter (12 bits)\\
 & &  and Extended Level 1 ID (32 bits)\\ \hline
\end{tabular}
\caption{Level 1 data and link count for inputs to the RoIB.}
\label{LVL1-data}}
 
The RoIB for the ATLAS HLT is a VME based system.  Assembled Level 1 data is  passed via S-Link to the Level 2 Supervisor Farm\cite{slink}.  The system is comprised of input cards and builder cards.  These cards are connected via a backplane that passes fragments from the input cards to the builder cards and passes flow control signals from the builder to the input cards.  An overview of the system appears in Fig.~\ref{LVL1-RoIB}.  The Level 1 data received are outlined in table~\ref{LVL1-data} and the components of the Level 1 system that communicate with the RoIB are indicated in figure~\ref{L1}.

\section{System Design}

The VME based system is composed of two types of cards.  The input cards receive data from the Level 1 system and distribute the data to the builder cards which assemble full event data and send these data to the level 2 supervisors.  Both cards are designed with very flexible electronics allowing implementation of most of the logic in the onboard FPGA's.  The input cards have 4 Altera APEX 20K200E FPGA's and the builder cards have 5 APEX 20K200E FPGA's plus support chips to implement the required logic. The system is intended to occupy a single 9U VME crate.  This includes a single board computer for control plus up to 4 input cards and up to 4 builder cards.

The firmware for the ATLAS TTC LDC, ROI Input, and ROI Builder Cards was developed and compiled using Altera's Quartus II design entry software.  The FPGA's on these boards are all packaged in Fine-Line Ball Grid Array packages. 
Each of the FPGA's is configured from an onboard EEPROM when powered-up.  Each of the EEPROM's is written with a Quartus II generated programming file, and is programmed in an Altera PROM programmer.

\subsection{Input Cards}

\smallskip
\FIGURE[l]{\epsfig{file=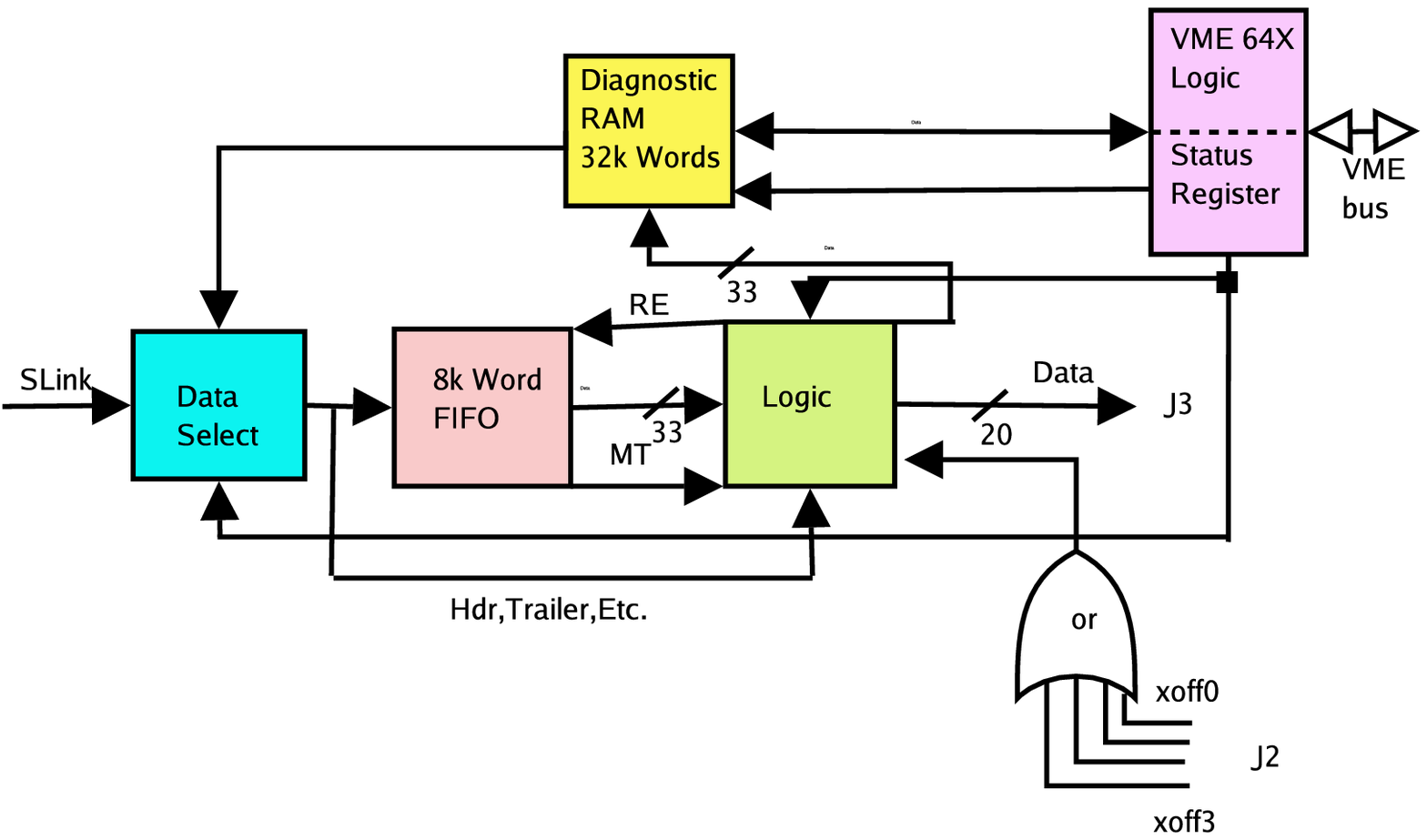,width=10cm} 
        \caption[Input Card Block Diagram]{A block diagram of one channel of the input card.}%
        \label{inputcard}}

The Region of Interest (RoI)  fragments are brought to the input cards of the RoIB via S-Link.  Each Input card accommodates 3 input S-Link mezzanine Link Destination Cards (LDCs) which receive and buffer the serial information from Level 1.  Each card can service up to 4 builder cards (limited by the custom backplane).  All transfer of information from the Input Cards to the builder cards is via J3 and a custom backplane mounted in the rear of the crate.  Each Input card also has a diagnostic RAM controlled from VMEbus which allows an onboard diagnostic system to emulate Level 1 fragments, and allows for verification of the RoIB system without the need for external inputs. The diagnostic RAMs on the input cards are 256K words deep.  The ROI Input Card uses four Altera APEX family EP20K200EFC484-3 FPGA's which are packaged in 484 BGA packages.

The Input Cards have several modes of operation.  {\it VME Mode} is used for initializing the Diagnostic RAM with diagnostic data and other control functions.  {\it Diagnostic Mode} is used to allow the RoIB system to perform its functions without any input from Level 1 trigger elements or from external devices emulating Level 1.  Input data streams are provided by the diagnostic RAMs resident on the Input Cards.  These RAMs are loaded from VMEbus in block transfers, and the contents may be data for diagnostic purposes (such as 5s and As, shifting 1s or 0s, etc.), or they may be test vectors from Monte Carlo simulation.  For these purposes, the S-Link output from the builder card could be routed via S-Link to a processor resident either in the crate or elsewhere which would verify or process the data.  
{\it Sniffer Mode} uses the diagnostic RAM to save fragments from the incoming Level 1 data streams.  The contents of the diagnostic RAM can then be accessed by block transfers initiated via VMEbus requests.  The received Fragments can be examined for diagnostic or system monitoring purposes.

Data from an Input Card are transferred through the custom backplane to the builder cards via J3, with the 3 input channels functioning independently of each other.  These incoming S-Link data includes both control and data words.  One channel is transferred on Row a, one on Row b, and one on Row c.  Each word is transferred in 2 twenty bit pieces, one after the other on a 20 MHz clock.  The first of the two 20 bit words consists of  the lower 16 bits of the data words, enable bit which is active for valid data, control bit which is always inactive, top word which is always inactive, and clock.  The second of the two 20 bit words consists of the upper 16 bits of the data words, enable bit which is always active for valid data, control bit which is active for control words and inactive for data words, top word which is always active, and clock.  The data are passed from each of the 3 channels of the input card via the custom backplane to the builder cards in parallel and received by each of them where the half words are concatenated and the fragment is reconstructed.  Because the system is capable of accommodating 12 input S-Link channels high density 250 pin connectors were used on J3 of the RoIB Cards, with the 10 extra pins tied to ground.

The flow control signals are transferred via pins on J2.  Since each builder card deals with input from 12 S-Link channels, there are 12 flow control signals.  These 12 signal lines are bussed on J2 and are wire-ored.  The first three of the flow control signals go to the first Input Card, the second three go to the second Input Card, etc. A simplified block diagram of the Input Card is shown in Fig.~\ref{inputcard}. There is an FPGA to handle the transactions with VME, which include 32 bit non-privileged transfers for reading/writing registers and 64 bit block transfers for reading/writing the diagnostic RAM(S).  The VME FPGA also includes a number of registers, such as status, which are relevant to all three channels.  The Input Card supports three input S-Link LDCs which buffer and deserialize RoI fragments from Level 1.  Each channel has an FPGA to provide the logic required for managing the data.  Depending on the Mode that is defined by the status register, there are two paths that data may take.  In both {\it Sniffer} and {\it Diagnostic mode} data passes from the FIFO through the FPGA.  If there are data in the FIFO and there is no Flow Control from the builder cards for the channel data are read from the FIFO one word at a time, parsed and formatted into 20 bit words, and transferred to the builder cards via J3 on a 20 MHz clock.  In {\it Sniffer Mode} the S-Link words are written to the Diagnostic RAM as they are read from FIFO, and in {\it Diagnostic Mode} the S-Link words are not written to RAM.  If the Diagnostic RAM has been written in {\it Sniffer Mode} it may subsequently be read from VME via 64 bit block transfers.

In {\it Diagnostic Mode}, data which have previously been written from VME to the Diagnostic RAM may be used to emulate real ROI Fragments from Level 1.  In this case the path of data is from RAM, through the FPGA, into the FIFO, from the FIFO, through the FPGA, and is then sent, through J3, to the builder cards.  In this mode the RAM is used as a source of input data and is not available for recording the input fragments.

\subsection{Builder Card}

\smallskip
\FIGURE[l]{\epsfig{file=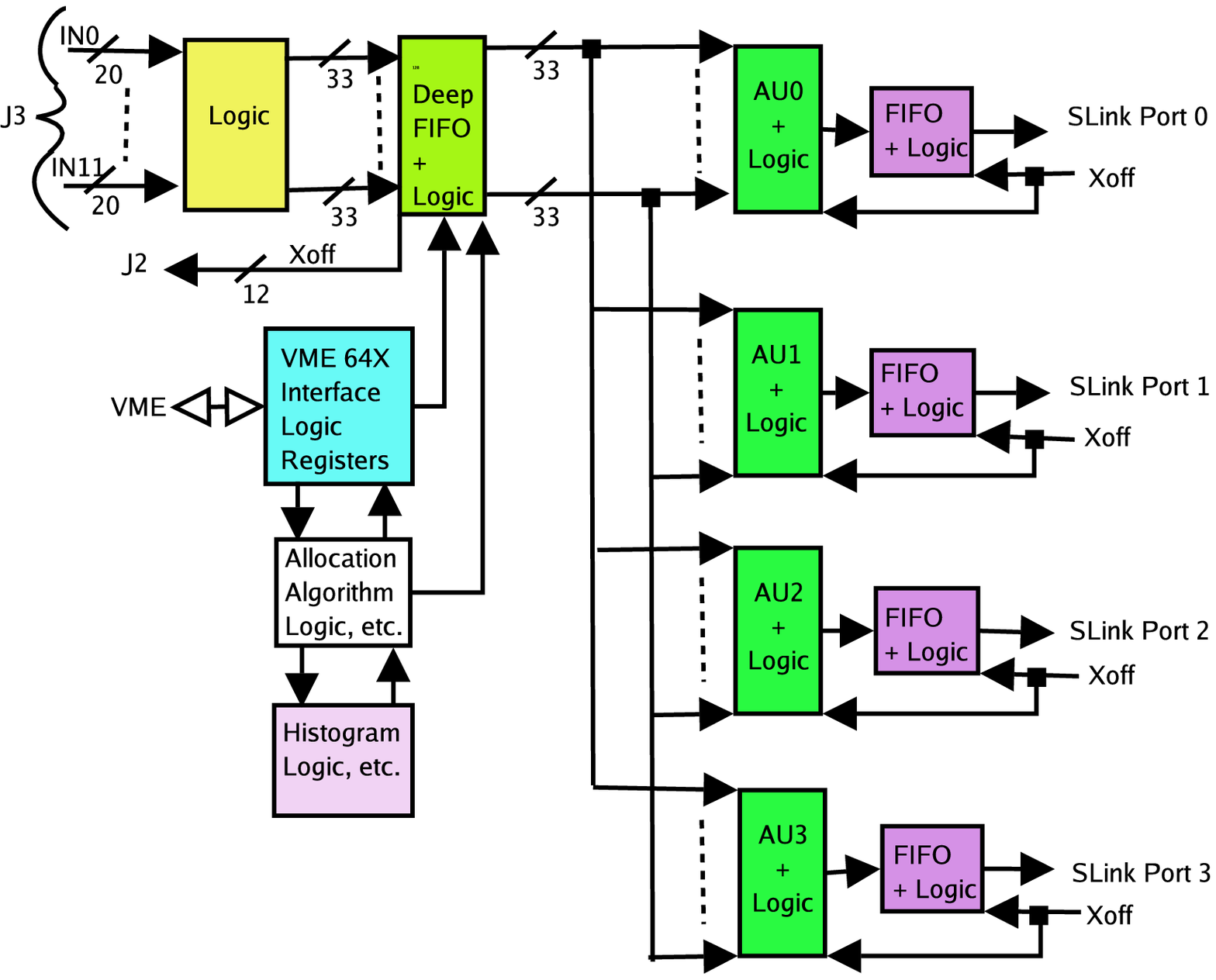,width=10cm} 
        \caption[Builder Card Block Diagram]{A block diagram of the builder card.}%
        \label{builder}}

We refer to the collected ROI fragments for a given event as an ROI record, and to the subsystem on the builder cards that builds the record, as the Assembly Unit (AU).  There are four AUs on each builder card.  The input cards pass ROI fragments to the builder cards.  Each builder card communicates ROI records via S-Link to four Supervisor processors (L2SV).   Each of the builder cards is responsible for a subset of the events that trigger Level 1.  Figure~\ref{builder} shows a simplified block diagram of the Builder card.
In the builder there is a basic round robin algorithm.  The system is expandable in units of four Supervisor processors by adding another builder card.  The backplane is able to accommodate four builder cards.  Each card has registers which tell it which of the Level 1 channels are active, how many builder cards there are and which card it is in the ordering.  The builder card uses five Altera APEX family EP20K200EFC484-3 FPGA's and one Altera APEX EP20K100FC484-3 FPGA which are all packaged in 484 BGA packages.
 
The event allocation algorithm must treat flow control and must deal with timeouts.  A timeout may occur as the result of a tardy fragment or a missing fragment.  The logic must distinguish the cases, and deal with either case  The events are allocated to Supervisor processors on a round robin basis, with the hardware dealing automatically with the number of cards, number of Supervisor processors, etc.  This scheme uses a 32 bit unique ID for each event constructed from a 24 bit Event Counter plus 8 higher order bits which count the number of times the Event Counter has been reset during the run.  This ID is referred to as the extended Level 1 ID, EL1ID.  The EL1ID is used throughout the system to identify which data belongs to which event.
The firmware also allows the allocation of events to specific AUs based on criteria other than the EL1ID. For example the Event Type could be used as the key for the selection of a builder card and AU on that card.

Each builder card is more or less autonomous.  For the testing described here the firmware allocated events to builder cards on the basis of {\it mod(EL1ID,number of active builder cards)}.  In every case the timeout system must interact with the event selection algorithm so that if a fragment is missing the event is handled properly.

\subsection{TTC/LDC Cards}

\smallskip
\FIGURE[l]{\epsfig{file=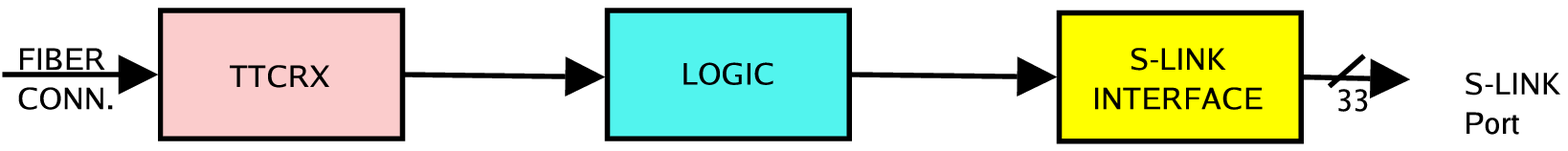,width=10cm} 
        \caption[TTC Mezzanine Block Diagram]{A block diagram of the TTC mezzanine card.}%
        \label{TTCmezzanine}}

The ATLAS experiment, like other LHC experiments, uses high speed control and timing system referred to as Trigger and Timing Control (TTC) to distribute the level 1 trigger decision to the front end systems.  The RoIB receives all of the information needed by the HLT via S-Link from components of the level 1 trigger, but as a cross check the RoIB provides a TTC input to the HLT.  This input was accomplished by designing a mezzanine card that had the same card connection as the S-Link LDC but received the TTC information directly via a front panel connection. Fig.~\ref{TTCmezzanine} shows a simple block diagram of the TTC LDC.  The TTC LDC uses one Altera ACEX family EP1K50FC256-3 FPGA which is packaged in a 256 BGA package.

This card receives the TTC fiber into a TTCRX chip which is a custom chip used by the control and experimental groups at the LHC.  The TTC information is latched on every L1A, and is formatted on the mezzanine card to resemble an ROI fragment, and is accepted and processed by the RoIB as if it were an ROI fragment.  The data which are latched on a L1A is the 24 bits of Event Counter, 12 bits of Bunch Counter, and 8 bits of Trigger Type.  When the data have been latched, they are reformatted in the form of an ROI Fragment, and written to a FIFO configured in the FPGA EABs.  If there is no Flow Control, the pseudo fragment is then transferred to the Input Card through the S-Link port.  An 8 bit register internally in the FPGA is incremented on Event Resets to provide the top 8 bits of the Level One Event number.  This is the same method used by the front end electronics to establish a full 32 bit EL1ID.

\subsection{Clock Card}
 
The ATLAS ROI Clock Card has a 20 MHz crystal oscillator that has its output buffered and driven out on 4 separate lines on the J2 plug-in backplane (one line to each of the 4 ROI Input Cards).  This is done to assure that the data from the different ROI Input Cards are all in phase.

\subsection{Flow Control}

A design constraint was that data integrity not be affected by flow control assertion at any point in the system.  There are deep FIFO's on every input so that the peaks of data activity are averaged, but flow control must propagate back via the individual S-Link channels to the Level 1 sources.  Flow control exists at a number of points in the system, and is not treated as one continuous signal.  For example, when a fragment is received on the builder in the input buffer FIFO for that particular input, Flow Control is activated by each builder in the crate back to the input card which provided that input.

The allocation algorithm examines the EL1ID and determines if the fragment is to be built in a record on this card.  If not, Flow Control to the input card is released.  If the allocation algorithm determines that this is the proper card, it then determines which of the four AU's on the card should build this record, and the logic begins to shift the fragment into that AU FIFO assuming it is not currently building a record.  When the trailer has been received by the AU FIFO the Flow Control back to the input card is released.  Because Flow Control is wire-ored on the backplane the input card continues to see Flow Control for this particular input until the fragment is contained in the appropriate AU FIFO.  At that time another fragment corresponding to the same input can be shifted to the builder cards.  If the AU selected by the allocation algorithm is currently building a previous record and the input buffer FIFO cannot shift its fragment into the AU FIFO, Flow Control is exerted back to the input card.  This input channel is then stopped and can not proceed until the AU FIFO is empty, which will not happen until the previous record times out or is completed and shifted out to the output FIFO.  This mechanism enforces a rate limit on the input to the HLT.

Another example of Flow Control on the RoIB is when the output FIFO goes half full, it exerts flow control back to the AU logic.  This typically happens while a record is being transferred into the output FIFO, and in that case the current record will be transferred in its entirety, but no other records will be transferred to that output FIFO until it is no longer half full.  An output FIFO begins transferring data to its S-Link port as soon as it is non-empty assuming Flow Control is not active on the S-Link.  If the output FIFO is half-full then  the L2SV is not servicing the S-Link input fast enough.  A possible reason for this is that the L2SV has crashed or some element of the system is unable to service the L1A rate.  If Flow Control is being exerted by the output S-Link and the output FIFO for that output is half full, no further records will be allocated to that L2SV. 

The individual ROI fragments can be as long as 128 S-Link words including headers and trailers, and are in the S-Link format.  This length constraint is imposed by the available EAB resources in the 20K200E FPGAs.  It is necessary to accommodate the time skew of arriving fragments, and accordingly a timer is started at the arrival of the first fragment of each event.  If all the fragments have been received before the timeout the compiled record will be transferred to the L2SV.  If the timeout occurs first the system transfers an incomplete record to the target L2SV.  The timeout and other parameters are of course selectable via a VMEbus transaction.  The maximum value of timeout that the system can implement is a critical parameter.  To the extent that a partially built ROI record has to wait for fragments, the RoIB card must provide buffering so that other records can be built concurrently.  The RoIB will  accommodate a timeout as long as 1 ms. A long timeout, however, places a strain on the system resources.  

If a fragment is lost, and an incomplete record is built, the event will be processed as is.  If a fragment is tardy and results in a timeout an incomplete record is built  In this case the tardy fragment must be recognized and discarded when it does arrive.  As far as the implementation in hardware is concerned, it is much easier to use local information to reject these tardy fragments.  This problem is dealt with by having the local allocation algorithm logic retain the EL1ID of the last complete record.  The local logic knows if it has built an incomplete record, and how many incomplete records there have been since the last complete record, and so knows what EL1ID subsequent fragments must have to be valid.  If a tardy fragment from a previous incompletely built record is received it is discarded.

In the event that a fragment is received without a header, that fragment must be ignored and treated as a missing fragment.  Since the allocation algorithm uses the EL1ID as the basic data input, and the position in the frame of the fragment of the EL1ID is determined in relation to the header, the allocation algorithm cannot reliably function, so the fragment must be ignored.    If the trailer is missing, the partial fragment will be built into the record.  The L2SV or other downstream trigger elements can determine that the fragment had an error by comparing fragment byte counts recorded in the record.  The input buffer on the builder card can contain 128 S-Link words.  In the event that a fragment is longer than 128 words the buffer will go full and the fragment will be truncated at the 128th word.  This truncation can also be detected by the L2SV or other trigger elements, but will not affect the building of the relevant record.

\section{Fabrication and Testing}

A complete system was built and tested to demonstrate that the RoIB would satisfy the ATLAS requirements.  Several small systems were built to provide test setups and to allow for software testing and software prototyping.  Tests were performed using a mixture of diagnostic RAM and real inputs prior to shipping the system to CERN.  The system installed for the ATLAS detector was tested with a complete set of 12 S-Link inputs.

\subsection{PC Board Fabrication and Testing}

All of the printed circuit boards' fabrication artwork was developed at Argonne National Laboratory using Cadence's Allegro PC Design Tools. 
The TTC LDC Card's dimensions are: 5.866" x 2.912" x 0.062".  There are 8 layers in the board, and it has 690 holes of which there are 13 different diameter sizes and the smallest diameter used is 0.010".  The smallest trace line width and line spacing is 0.006".  The board material is FR-4, and the finished board has an Electroless-Nickel Immersion Gold (ENIG) finish.  It has soldermask and legends on both sides. 

The ROI Input Card's dimensions are: 15.750" x 14.437" x 0.093".  The top and bottom edges are milled down to 0.062" thickness to accommodate the VME crate card guides.  There are 12 layers in the board, and it has 5685 holes of which there are 13 different diameter sizes and the smallest diameter used is 0.010".  The smallest trace line width and line spacing is 0.006".  The board material is FR-4, and the finished board has an Electroless-Nickel Immersion Gold (ENIG) finish.  It has soldermask and legends on both sides.

The ROI Builder Card's dimensions are: 15.750" x 14.437" x 0.093".  The top and bottom edges are milled down to 0.062" thickness to accommodate the VME crate card guides.  There are 18 layers in the board, and it has 8731 holes of which there are 12 different diameter sizes and the smallest diameter used is 0.010".  The smallest trace line width and line spacing is 0.006".  The board material is FR-4, and the finished board has a Hot Air Solder Leveling (HASL) finish.  It has soldermask and legends on both sides.

The ROI Clock Card's dimensions are: 15.750" x 14.437" x 0.062".  There are 4 layers in the board, and it has 365 holes of which there are 5 different diameter sizes and the smallest diameter used is 0.029".  The smallest trace line width and line spacing is 0.010".  The board material is FR-4, and the finished board has a Hot Air Solder Leveling (HASL) finish.  It has soldermask on both sides and the legend on the top side.
 
The 9U x 400mm VME ROI Input, Builder, and Clock cards all have stiffener bars to help keep the boards flat.  They also have Injector/Ejector levers to aid in insertion and extraction of the cards in the VME crate.
 
The J2 plug-in backplane dimensions are: 7.200" x 3.900" x 0.125".  There are 6 layers in the board, and it has 1474 holes of which there are 3 different hole sizes and the smallest diameter used is 0.015".  The smallest trace line width and line spacing is 0.010".  The board material is FR-4, and the finished board has a Hot Air Solder Leveling (HASL) finish.  It has soldermask and legends on both sides.
 
The J3 backplane's dimensions are 8.700" x 5.067" x 0.125".  There are 14 layers in the board, and it has 3206 holes of which there are 3 different hole sizes and the smallest diameter used is 0.025".  The smallest trace line width and line spacing is 0.010".  The board material is FR-4, and the finished board has a Hot Air Solder Leveling (HASL) finish.  It has soldermask and legends on both sides.  This backplane has surface-mount pads for terminating the data lines.

Each ROI Input Card has six sets of three 74AHC16541 buffer IC's soldered in
parallel to transmit the data to the ROI Builder Cards.  Each ROI Builder
Card receives the data from the Input Cards with fifteen 74LVCH16541A buffer
IC's.

The J3 backplane was designed with a characteristic line impedance of 100
ohms.  There is space on this backplane for onboard terminations but stable operation did not require these so they were not added.

Cards were tested in a small setup using a combination of external inputs and diagnostic memory.  Special standalone programs were used to send, receive and test the data integrity.

\subsection{System Testing}

Tests were performed to evaluate the rate capability, flow control and the correct handling of missing fragments from the Level 1 input.  Software to allow for testing in standalone and test setups plus under the control of the ATLAS online system was developed and used for all these tests.

In order to test assembly of the fragments and to demonstrate correct performance with flow control on some of the outputs,  tests were run with multiple outputs but only a subset of the output channels were checked for correct results.  In one test a single output channel was monitored while running the others without checking at the maximum rate possible.  Figure~\ref{Rate-test} shows the performance of this configuration where a single L2SV (the one that checked the results for that output) dictated the system speed by asserting flow control.  The outputs that were not checked were configured to ignore the data and never assert flow control.  For a typical fragment size of 18 words this shows that a system with 8 inputs and 8 outputs would run at 320kHz Level 1 input if the routine used to check the single output is typical of the rate for the event dispatching function performed by the L2SV in the ATLAS experiment.

\FIGURE[l]{\epsfig{file=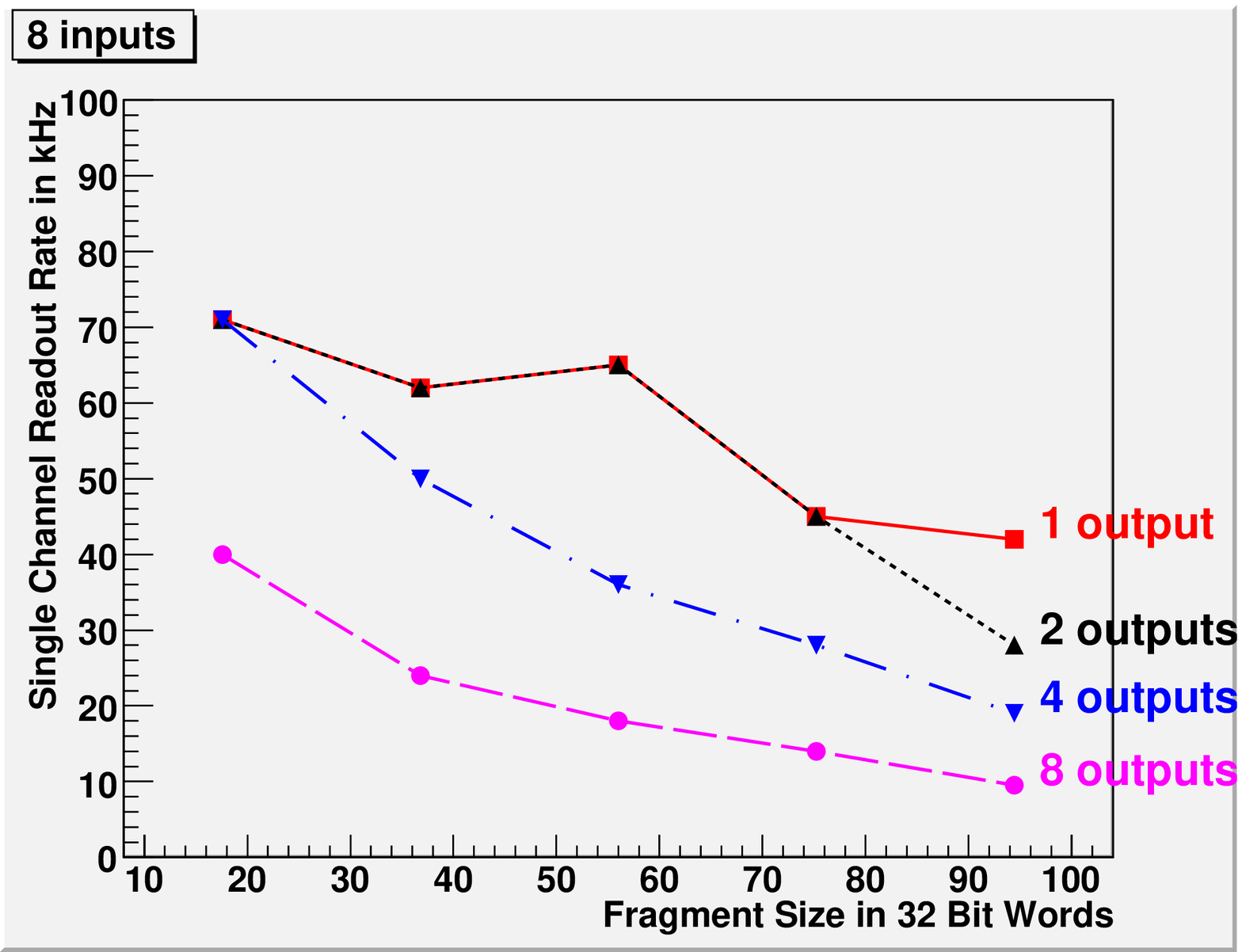,width=10cm} 
        \caption[Rate test]{The rate to each L2SV for event data as a function of the size of the Level 1 fragments.  These tests were done with 8 inputs from diagnostic RAM.  Each curve was taken with a different number of outputs (L2SVs).}%
        \label{Rate-test}}

A second test was performed that varied the number of outputs while checking 4 output channels (all readout by the same FILAR).  The system reading the data out was an SMP dual 2.4GHz, XEON with hyperthreading enabled so there were in effect 4 threads available.  The input did not fully take advantage of the multithreading but the testing was done in separate threads.  Again the other outputs were setup to never assert flow control.  Varying size fragments were tested as well as the case where all the fragment sizes were the same.  The results of this latter case are shown in Figure~\ref{Check-test}.

\FIGURE[l]{\epsfig{file=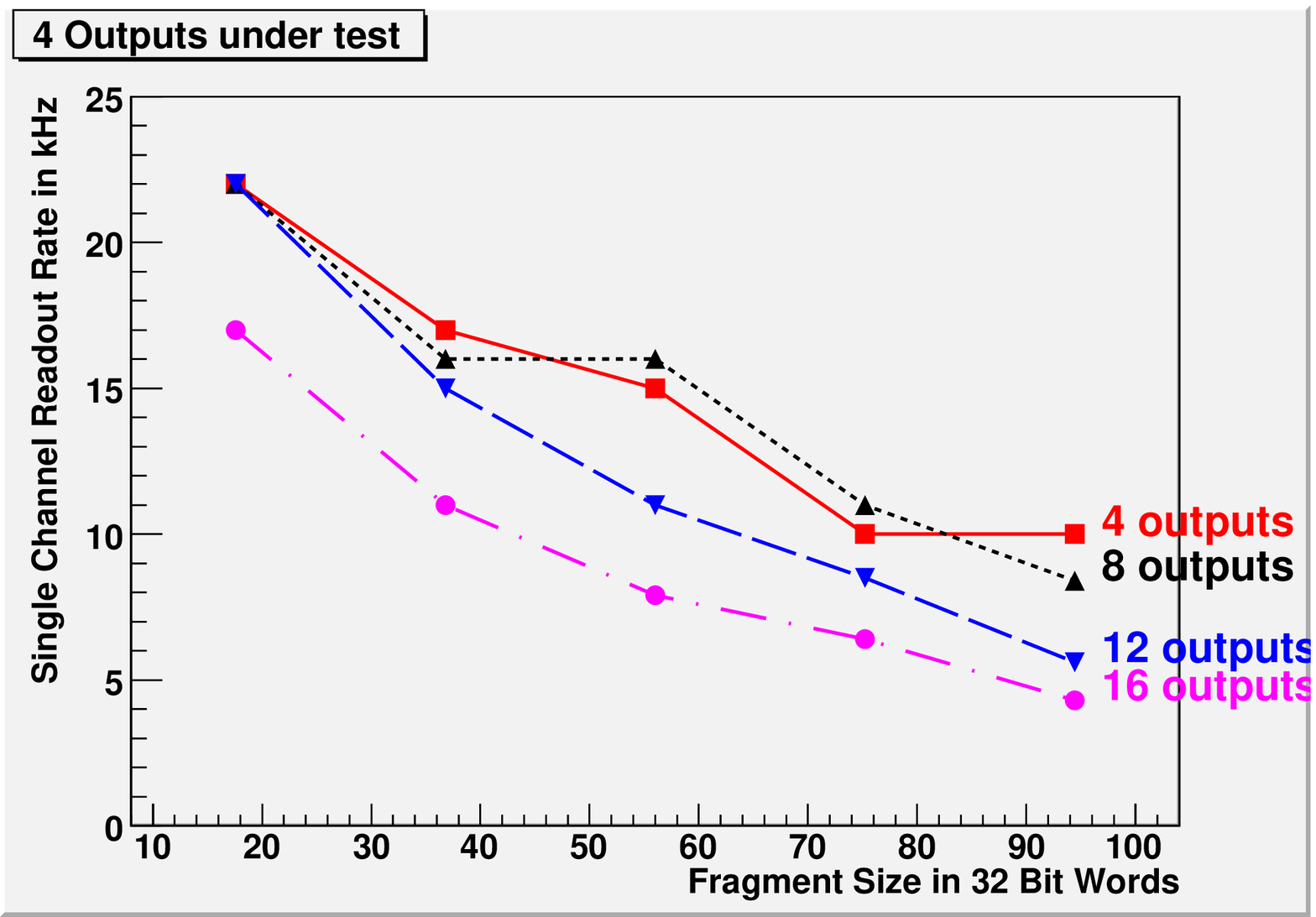,width=10cm} 
        \caption[Multiple checking test]{The rate to each L2SV for event data as a function of the size of the Level 1 fragments.  These tests were done with 8 inputs from diagnostic RAM.  Each curve was taken with a different number of outputs (L2SVs).  In all cases 4 outputs were checked and the remaining asserted no flow control.}%
        \label{Check-test}}

  Tests performed after the system was installed using an external input source were limited by the external source's rate capability.  The system performed fragment assembly with 8 inputs using fragments from ATLAS simulations at 115 kHz.  For these tests no flow control back from the L2SVs was asserted so the ability of the L2SVs to handle the rate was not a limiting factor.  This shows that the rate limit will be established by the assertion of flow control and the speed of the HLT plus L2SV software and not by the RoIB hardware.

\end{document}